
\input phyzzx.tex
\catcode`\@=11
\def\wlog#1{}
\def\eqname#1{\rel@x {\pr@tect
  \ifnum\equanumber<0 \xdef#1{{\rm\number-\equanumber}}%
     \gl@bal\advance\equanumber by -1
  \else \gl@bal\advance\equanumber by 1
     \ifx\chapterlabel\rel@x \def\d@t{}\else \def\d@t{.}\fi
    \xdef#1{{\rm\chapterlabel\d@t\number\equanumber}}\fi #1}}
\catcode`\@=12
\catcode`\@=11

\def\eat@#1{}
\mathchardef\prime@="0230
\def\prime{{{}\prime@{}}}
\def\prim@s{\prime@\futurelet\next\pr@m@s}

\def\,{\relax\ifmmode\mskip\thinmuskip\else\thinspace\fi}
\def\!{\relax\ifmmode\mskip-\thinmuskip\else\negthinspace\fi}
\def\frac#1#2{{#1\over#2}}
\def\dfrac#1#2{{\displaystyle{#1\over#2}}}

\def\:{\nobreak\hskip.1111em{:}\hskip.3333em plus .0555em\relax}
\def\intic@{\mathchoice{\hskip5\p@}{\hskip4\p@}{\hskip4\p@}{\hskip4\p@}}
\def\negintic@
 {\mathchoice{\hskip-5\p@}{\hskip-4\p@}{\hskip-4\p@}{\hskip-4\p@}}
\def\intkern@{\mathchoice{\!\!\!}{\!\!}{\!\!}{\!\!}}
\def\intdots@{\mathchoice{\cdots}{{\cdotp}\mkern1.5mu
    {\cdotp}\mkern1.5mu{\cdotp}}{{\cdotp}\mkern1mu{\cdotp}\mkern1mu
      {\cdotp}}{{\cdotp}\mkern1mu{\cdotp}\mkern1mu{\cdotp}}}
\newcount\intno@
\def\iint{\intno@=\tw@\futurelet\next\ints@}
\def\iiint{\intno@=\thr@@\futurelet\next\ints@}
\def\iiiint{\intno@=4 \futurelet\next\ints@}
\def\idotsint{\intno@=\z@\futurelet\next\ints@}
\def\ints@{\findlimits@\ints@@}
\newif\iflimtoken@
\newif\iflimits@
\def\findlimits@{\limtoken@false\limits@false\ifx\next\limits
 \limtoken@true\limits@true
   \else\ifx\next\nolimits\limtoken@true\limits@false
    \fi\fi}
\def\multintlimits@{\intop\ifnum\intno@=\z@\intdots@
  \else\intkern@\fi
    \ifnum\intno@>\tw@\intop\intkern@\fi
     \ifnum\intno@>\thr@@\intop\intkern@\fi\intop}
\def\multint@{\int\ifnum\intno@=\z@\intdots@\else\intkern@\fi
   \ifnum\intno@>\tw@\int\intkern@\fi
    \ifnum\intno@>\thr@@\int\intkern@\fi\int}
\def\ints@@{\iflimtoken@\def\ints@@@{\iflimits@
   \negintic@\mathop{\intic@\multintlimits@}\limits\else
    \multint@\nolimits\fi\eat@}\else
     \def\ints@@@{\multint@\nolimits}\fi\ints@@@}
\def\Sb{_\bgroup\vspace@
        \baselineskip=\fontdimen10 \scriptfont\tw@
        \advance\baselineskip by \fontdimen12 \scriptfont\tw@
        \lineskip=\thr@@\fontdimen8 \scriptfont\thr@@
        \lineskiplimit=\thr@@\fontdimen8 \scriptfont\thr@@
        \Let@\vbox\bgroup\halign\bgroup \hfil$\scriptstyle
            {##}$\hfil\cr}
\def\endSb{\crcr\egroup\egroup\egroup}
\def\Sp{^\bgroup\vspace@
        \baselineskip=\fontdimen10 \scriptfont\tw@
        \advance\baselineskip by \fontdimen12 \scriptfont\tw@
        \lineskip=\thr@@\fontdimen8 \scriptfont\thr@@
        \lineskiplimit=\thr@@\fontdimen8 \scriptfont\thr@@
        \Let@\vbox\bgroup\halign\bgroup \hfil$\scriptstyle
            {##}$\hfil\cr}
\def\endSp{\crcr\egroup\egroup\egroup}
\def\Let@{\relax\iffalse{\fi\let\\=\cr\iffalse}\fi}
\def\vspace@{\def\vspace##1{\noalign{\vskip##1 }}}
\def\aligned{\,\vcenter\bgroup\vspace@\Let@\openup\jot\m@th\ialign
  \bgroup \strut\hfil$\displaystyle{##}$&$\displaystyle{{}##}$\hfil\crcr}
\def\endaligned{\crcr\egroup\egroup}
\def\matrix{\,\vcenter\bgroup\Let@\vspace@
    \normalbaselines
  \m@th\ialign\bgroup\hfil$##$\hfil&&\quad\hfil$##$\hfil\crcr
    \mathstrut\crcr\noalign{\kern-\baselineskip}}
\def\endmatrix{\crcr\mathstrut\crcr\noalign{\kern-\baselineskip}\egroup
                \egroup\,}
\newtoks\hashtoks@
\hashtoks@={#}
\def\format{\crcr\egroup\iffalse{\fi\ifnum`}=0 \fi\format@}
\def\format@#1\\{\def\preamble@{#1}%
  \def\c{\hfil$\the\hashtoks@$\hfil}%
  \def\r{\hfil$\the\hashtoks@$}%
  \def\l{$\the\hashtoks@$\hfil}%
  \setbox\z@=\hbox{\xdef\Preamble@{\preamble@}}\ifnum`{=0 \fi\iffalse}\fi
   \ialign\bgroup\span\Preamble@\crcr}

\def\cases{\left\{\,\vcenter\bgroup\vspace@
     \normalbaselines\openup\jot\m@th
       \Let@\ialign\bgroup$##$\hfil&\quad$##$\hfil\crcr
      \mathstrut\crcr\noalign{\kern-\baselineskip}}

\newif\iftagsleft@
\tagsleft@true
\def\TagsOnRight{\global\tagsleft@false}
\def\tag#1$${\iftagsleft@\leqno\else\eqno\fi
 \hbox{\def\pagebreak{\global\postdisplaypenalty-\@M}%
 \def\nopagebreak{\global\postdisplaypenalty\@M}\rm(#1\unskip)}%
  $$\postdisplaypenalty\z@\ignorespaces}
\interdisplaylinepenalty=\@M
\def\allowdisplaybreak@{\def\allowdisplaybreak{\noalign{\allowbreak}}}
\def\displaybreak@{\def\displaybreak{\noalign{\break}}}
\def\align#1\endalign{\def\tag{&}\vspace@\allowdisplaybreak@\displaybreak@
  \iftagsleft@\lalign@#1\endalign\else
   \ralign@#1\endalign\fi}
\def\ralign@#1\endalign{\displ@y\Let@\tabskip\centering
   \halign to\displaywidth
     {\hfil$\displaystyle{##}$\tabskip=\z@&$\displaystyle{{}##}$\hfil
       \tabskip=\centering&\llap{\hbox{(\rm##\unskip)}}\tabskip\z@\crcr
             #1\crcr}}
\def\lalign@
 #1\endalign{\displ@y\Let@\tabskip\centering\halign to \displaywidth
   {\hfil$\displaystyle{##}$\tabskip=\z@&$\displaystyle{{}##}$\hfil
   \tabskip=\centering&\kern-\displaywidth
        \rlap{\hbox{(\rm##\unskip)}}\tabskip=\displaywidth\crcr
               #1\crcr}}
\def\overrightarrow{\mathpalette\overrightarrow@}
\def\overrightarrow@#1#2{\vbox{\ialign{$##$\cr
    #1{-}\mkern-6mu\cleaders\hbox{$#1\mkern-2mu{-}\mkern-2mu$}\hfill
     \mkern-6mu{\to}\cr
     \noalign{\kern -1\p@\nointerlineskip}
     \hfil#1#2\hfil\cr}}}
\def\overleftarrow{\mathpalette\overleftarrow@}
\def\overleftarrow@#1#2{\vbox{\ialign{$##$\cr
     #1{\leftarrow}\mkern-6mu\cleaders
      \hbox{$#1\mkern-2mu{-}\mkern-2mu$}\hfill
      \mkern-6mu{-}\cr
     \noalign{\kern -1\p@\nointerlineskip}
     \hfil#1#2\hfil\cr}}}
\def\overleftrightarrow{\mathpalette\overleftrightarrow@}
\def\overleftrightarrow@#1#2{\vbox{\ialign{$##$\cr
     #1{\leftarrow}\mkern-6mu\cleaders
       \hbox{$#1\mkern-2mu{-}\mkern-2mu$}\hfill
       \mkern-6mu{\to}\cr
    \noalign{\kern -1\p@\nointerlineskip}
      \hfil#1#2\hfil\cr}}}
\def\underrightarrow{\mathpalette\underrightarrow@}
\def\underrightarrow@#1#2{\vtop{\ialign{$##$\cr
    \hfil#1#2\hfil\cr
     \noalign{\kern -1\p@\nointerlineskip}
    #1{-}\mkern-6mu\cleaders\hbox{$#1\mkern-2mu{-}\mkern-2mu$}\hfill
     \mkern-6mu{\to}\cr}}}
\def\underleftarrow{\mathpalette\underleftarrow@}
\def\underleftarrow@#1#2{\vtop{\ialign{$##$\cr
     \hfil#1#2\hfil\cr
     \noalign{\kern -1\p@\nointerlineskip}
     #1{\leftarrow}\mkern-6mu\cleaders
      \hbox{$#1\mkern-2mu{-}\mkern-2mu$}\hfill
      \mkern-6mu{-}\cr}}}
\def\underleftrightarrow{\mathpalette\underleftrightarrow@}
\def\underleftrightarrow@#1#2{\vtop{\ialign{$##$\cr
      \hfil#1#2\hfil\cr
    \noalign{\kern -1\p@\nointerlineskip}
     #1{\leftarrow}\mkern-6mu\cleaders
       \hbox{$#1\mkern-2mu{-}\mkern-2mu$}\hfill
       \mkern-6mu{\to}\cr}}}
\def\sqrt#1{\radical"270370 {#1}}
\def\dots{\relax\ifmmode\let\next=\ldots\else\let\next=\tdots@\fi\next}
\def\tdots@{\unskip\ \tdots@@}
\def\tdots@@{\futurelet\next\tdots@@@}
\def\tdots@@@{$\mathinner{\ldotp\ldotp\ldotp}\,
   \ifx\next,$\else
   \ifx\next.\,$\else
   \ifx\next;\,$\else
   \ifx\next:\,$\else
   \ifx\next?\,$\else
   \ifx\next!\,$\else
   $ \fi\fi\fi\fi\fi\fi}
\def\text{\relax\ifmmode\let\next=\text@\else\let\next=\text@@\fi\next}
\def\text@@#1{\hbox{#1}}
\def\text@#1{\mathchoice
 {\hbox{\everymath{\displaystyle}\def\textfonti{\the\textfont1 }%
    \def\textfontii{\the\textfont2 }\textdef@@ T#1}}
 {\hbox{\everymath{\textstyle}\def\textfonti{\the\textfont1 }%
    \def\textfontii{\the\textfont2 }\textdef@@ T#1}}
 {\hbox{\everymath{\scriptstyle}\def\textfonti{\the\scriptfont1 }%
   \def\textfontii{\the\scriptfont2 }\textdef@@ S\rm#1}}
 {\hbox{\everymath{\scriptscriptstyle}%
   \def\textfonti{\the\scriptscriptfont1 }%
   \def\textfontii{\the\scriptscriptfont2 }\textdef@@ s\rm#1}}}
\def\textdef@@#1{\textdef@#1\rm \textdef@#1\bf
   \textdef@#1\sl \textdef@#1\it}

\def\textdef@#1#2{%
 \def\next{\csname\expandafter\eat@\string#2fam\endcsname}%
\if S#1\edef#2{\the\scriptfont\next\relax}%
 \else\if s#1\edef#2{\the\scriptscriptfont\next\relax}%
 \else\edef#2{\the\textfont\next\relax}\fi\fi}
\scriptfont\itfam=\tenit \scriptscriptfont\itfam=\tenit
\scriptfont\slfam=\tensl \scriptscriptfont\slfam=\tensl
\mathcode`\0="0030
\mathcode`\1="0031
\mathcode`\2="0032
\mathcode`\3="0033
\mathcode`\4="0034
\mathcode`\5="0035
\mathcode`\6="0036
\mathcode`\7="0037
\mathcode`\8="0038
\mathcode`\9="0039
\def\Cal{\relax\ifmmode\let\next=\Cal@\else
    \def\next{\errmessage{Use \string\Cal\space only in %
      math mode}}\fi\next}
    \def\Cal@#1{{\fam2 #1}}
\def\bold{\relax\ifmmode\let\next=\bold@\else
    \def\next{\errmessage{Use \string\bold\space only in %
      math mode}}\fi\next}
    \def\bold@#1{{\fam\bffam #1}}
\mathchardef\Gamma="0000
\mathchardef\Delta="0001
\mathchardef\Theta="0002
\mathchardef\Lambda="0003
\mathchardef\Xi="0004
\mathchardef\Pi="0005
\mathchardef\Sigma="0006
\mathchardef\Upsilon="0007
\mathchardef\Phi="0008
\mathchardef\Psi="0009
\mathchardef\Omega="000A
\mathchardef\varGamma="0100
\mathchardef\varDelta="0101
\mathchardef\varTheta="0102
\mathchardef\varLambda="0103
\mathchardef\varXi="0104
\mathchardef\varPi="0105
\mathchardef\varSigma="0106
\mathchardef\varUpsilon="0107
\mathchardef\varPhi="0108
\mathchardef\varPsi="0109
\mathchardef\varOmega="010A
\def\wlog#1{\immediate\write-1{#1}}
\catcode`\@=12  
\def\=def{\; \mathop{=}_{\text{\rm def}} \;}
\def\rd{\partial}
\def\Res{\mathop{\;\text{Res}\;}}
\def\bhat{{\hat{b}}}
\def\uhat{{\hat{u}}}
\def\vhat{{\hat{v}}}
\def\what{{\hat{w}}}
\def\zhat{{\hat{z}}}
\def\Bhat{{\hat{B}}}
\def\Hhat{{\hat{H}}}
\def\Lhat{{\hat{L}}}
\def\Mhat{{\hat{M}}}
\def\Shat{{\hat{S}}}
\def\What{{\hat{W}}}
\def\Zhat{{\hat{Z}}}

\def\bfZ{{\bold Z}}
\def\calB{{\cal B}}
\def\calL{{\cal L}}
\def\calM{{\cal M}}

\def\calBhat{{\hat{\calB}}}
\def\calLhat{{\hat{\calL}}}
\def\calMhat{{\hat{\calM}}}
\def\Psihat{{\hat{\Psi}}}
\def\rdtilde{{\tilde{\rd}}}
\def\lambdatilde{{\tilde{\lambda}}}
\def\lambdahat{{\hat{\lambda}}}

\hsize=15.5truecm
\vsize=23truecm
\sequentialequations
\doublespace
\TagsOnRight
\overfullrule=0pt
\sectionstyle={\Number}
\pubnum={Kyoto University KUCP-0057/93}
\date={January 1993}
\titlepage
\title{\fourteencp Quasi-classical limit of Toda hierarchy
  and W-infinity symmetries}
\author{Kanehisa Takasaki}
\address{
  Department of Fundamental Sciences\break
  Faculty of Integrated Human Studies, Kyoto University\break
  Yoshida-Nihonmatsu-cho, Sakyo-ku, Kyoto 606, Japan\break
  E-mail: takasaki @ jpnyitp (Bitnet)\break
}
\andauthor{Takashi Takebe}
\address{
  Department of Mathematical Sciences, University of Tokyo\break
  Hongo, Bunkyo-ku, Tokyo 113, Japan\break
  E-mail: takebe @ math.s.u-tokyo.ac.jp\break
}
\abstract
\noindent
Previous results on quasi-classical limit of the KP hierarchy and
its W-infinity symmetries are extended to the Toda hierarchy.
The Planck constant $\hbar$ now emerges as the spacing unit of
difference operators in the Lax formalism. Basic notions, such as
dressing operators, Baker-Akhiezer functions and tau function,
are redefined. $W_{1+\infty}$ symmetries of the Toda hierarchy
are realized by suitable rescaling of the Date-Jimbo-Kashiara-Miwa
vertex operators. These symmetries are contracted to $w_{1+\infty}$
symmetries of the dispersionless hierarchy through their action
on the tau function.

\endpage
\section{Introduction}

\noindent
Dispersionless analogues of integrable systems of KP and Toda
type
[\REF\displess{
  Lebedev, D., and Manin, Yu.,
  Conservation Laws and Lax Representation on Benny's Long Wave Equations,
  Phys.Lett. 74A (1979), 154--156. \nextline
  Kodama, Y.,
  A method for solving the dispersionless KP equation and
  its exact solutions,
  Phys. Lett. 129A (1988), 223-226;
  Solutions of the dispersionless Toda equation,
  Phys. Lett. 147A (1990), 477-482. \nextline
  Kodama, Y., and Gibbons, J.,
  A method for solving the dispersionless KP hierarchy and
  its exact solutions, II,
  Phys. Lett. 135A (1989), 167-170.}
\displess]
provide an interesting family of integrable ``contractions."
In the context of field theory, the dispersionless Toda equation
is studied as continuous (or large-$N$) limit of the ordinary
Toda field theory
[\REF\TodaFT{
  Bakas, I.,
   The structure of the $W_\infty$ algebra,
  Commun. Math. Phys. 134 (1990), 487-508.\nextline
  Saveliev, M.V., and Vershik, A.M.,
   Continual analogues of contragredient Lie algebras,
  Commun. Math. Phys. 126 (1989), 367-378.}
\TodaFT]
as well as a dimensional reduction of
four dimensional selfdual gravity
[\REF\SDG{
  Bakas, I.,
  Area preserving diffeomorphisms and
  higher spin fields in two dimensions,
  in {\it Supermembranes and Physics in 2+1 Dimensions},
  Trieste 1989, M. Duff, C. Pope and E. Sezgin eds.
  (World Scientific, 1990).\nextline
  Park, Q-Han,
  Extended conformal symmetries in real heavens,
  Phys. Lett. 236B (1990), 429-432.}
\SDG].
Recently a hierarchy of higher flows are constructed
[\REF\TaTadToda{
  Takasaki, K., and Takebe, T.,
   SDiff(2) Toda equation -- hierarchy, tau function and symmetries,
  Lett. Math. Phys. 23 (1991), 205-214.}
\TaTadToda]
and applied to two dimensional string theory
[\REF\AvJe{
  Avan, J.,
  $w_\infty$-currents in 3-dimensional Toda theory,
  BROWN-HET-855 (March, 1992).\nextline
  Avan, J., and Jevicki, A.,
  Interacting theory of collective and topological fields
  in 2 dimensions, BROWN-HET-869 (August, 1992).}
\AvJe].

Dispersionless (or long wavelength) limit can also be
understood as quasi-classical limit.  In a previous paper
[\REF\TaTaQCKP{
  Takasaki, K., and Takebe, T.,
  Quasi-classical limit of KP hierarchy, W-symmetries and free fermions,
  Kyoto preprint KUCP-0050/92 (July, 1992).}
\TaTaQCKP],
we considered the KP hierarchy and its dispersionless version from
this point of view, and could show a direct connection
between W-infinity symmetries of the two hierarchies, i.e.,
$W_{1+\infty}$ symmetries of the KP hierarchy and
$w_{1+\infty}$ symmetries of the dispersionless KP hierarchy.
In this note, we present similar results on the Toda hierarchy
and its dispersionless version.

\section{Lax formalism of Dispersionless Toda hierarchy}

\noindent
To begin with, let us briefly review the Lax formalism of the
dispersionless Toda hierarchy [\TaTadToda].
The dispersionless Toda hierarchy consists of an infinite number
of commuting flows with ``time variables" $z = (z_1,z_2,\ldots)$
and $\zhat = (\zhat_1,\zhat_2,\ldots)$. The Lax equations can be
written
$$\align
  \frac{\rd \calL}{\rd z_n} = \{ \calB_n, \calL \}, \quad&
  \frac{\rd \calL}{\rd \zhat_n} = \{ \calBhat_n, \calL \},
                                                                  \\
  \frac{\rd \calLhat}{\rd z_n} = \{ \calB_n, \calLhat \}, \quad&
  \frac{\rd \calLhat}{\rd \zhat_n} = \{ \calBhat_n, \calLhat \},
  \quad n = 1,2,\ldots,                                   \tag\eq \\
\endalign
$$
where $\calL$ and $\calLhat$ are Laurent series
$$
\align
  \calL =& p + \sum_{n=0}^\infty u_n(z,\zhat,s) p^{-n},          \\
  \calLhat =& \sum_{n=1}^\infty \uhat_n(z,\zhat,s) p^n    \tag\eq \\
\endalign
$$
of a variable $p$, and $\calB_n$ and $\calBhat_n$ are given by
$$
   \calB_n = (\calL^n)_{\ge 0}, \quad
   \calBhat_n = (\calLhat^{-n})_{\le -1}.                \tag\eq
$$
Here $(\quad)_{\ge 0}$ and $(\quad)_{\le -1}$ denote the projection
of Laurent series onto a linear combination of $p^n$ with
$n\ge 0$ and $\le -1$ respectively. The Poisson bracket
$\{\quad,\quad\}$ is defined by
$$
  \{ A(p,s), B(p,s) \}
  = p\frac{\rd A(p,s)}{\rd p} \frac{\rd B(p,s)}{\rd s}
   -\frac{\rd A(p,s)}{\rd s} p\frac{\rd B(p,s)}{\rd p}  \tag\eq
$$
on the two dimensional ``phase space" with coordinates $(p,s)$.
This Lax system can be extended to a larger system. The extended
Lax representation possesses, in addition to the above equations,
another set of dispersionless Lax equations
$$
\align
  \frac{\rd \calM}{\rd z_n} = \{ \calB_n, \calM \}, \quad&
  \frac{\rd \calM}{\rd \zhat_n} = \{ \calBhat_n, \calM \},
                                                                  \\
  \frac{\rd \calMhat}{\rd z_n} = \{ \calB_n, \calMhat \}, \quad&
  \frac{\rd \calMhat}{\rd \zhat_n} = \{ \calBhat_n, \calMhat \},
  \quad n = 1,2,\ldots,                                   \tag\eq \\
\endalign
$$
and the canonical Poisson relations
$$
  \{ \calL, \calM \} = \calL, \quad
  \{ \calLhat, \calMhat \} = \calLhat                     \tag\eq
$$
for a second set of Laurent series
$$
\align
  \calM =& \sum_{n=1}^\infty n z_n \calL^n + s
          +\sum_{n=1}^\infty v_n(z,\zhat,s)\calL^{-n},
                                                                   \\
  \calMhat =& -\sum_{n=1}^\infty n \zhat_n \calLhat^{-n} + s
              +\sum_{n=1}^\infty \vhat_n(z,\zhat,s)\calLhat^n.
                                                           \tag\eq \\
\endalign
$$
These somewhat complicated equations can actually be cast into a
simple, compact 2-form equation:
$$
    \frac{d\calL \wedge d\calM}{\calL}
    = \omega
    = \frac{d\calLhat \wedge d\calMhat}{\calLhat}
                                              \tag\eqname\TwoForm
$$
where
$$
  \omega  = \frac{dp}{p}\wedge ds
      + \sum_{n=1}^\infty d\calB_n \wedge dz_n
      + \sum_{n=1}^\infty d\calBhat_n \wedge d\zhat_n.   \tag\eq
$$

\section{Lax formalism of Toda hierarchy with Planck constant}

\noindent
To interpret this hierarchy as quasi-classical limit,
we reformulate the ordinary Toda hierarchy
[\REF\UeTa{
  Ueno, K., and Takasaki, K.,
  Toda lattice hierarchy,
  in {\it Group Representations and Systems of Differential Equations},
  K. Okamoto ed., Advanced Studies in Pure Math. 4
  (North-Holland/Kinokuniya 1984).\nextline
  Takasaki, K.,
  Initial value problem for the Toda lattice hierarchy,
  ibid.}
\UeTa]
in the language of
difference operators in an continuous variable $s$ with spacing unit
$\hbar$. The Lax equations are then given by
$$
\align
  \hbar \frac{\rd L}{\rd z_n} = [ B_n, L ], \quad&
  \hbar \frac{\rd L}{\rd \zhat_n} = [\Bhat_n, L ],
                                                                  \\
  \hbar \frac{\rd \Lhat}{\rd z_n} = [ B_n, \Lhat ], \quad&
  \hbar \frac{\rd \Lhat}{\rd \zhat_n} = [ \Bhat_n, \Lhat ],
  \quad n = 1,2,\ldots,                                   \tag\eq \\
\endalign
$$
where the Lax operators $L$ and $\Lhat$ are difference operators
of the form
$$
\align
  L =& e^{\hbar\rd/\rd s} + \sum_{n=0}^\infty
                   u_n(\hbar,z,\zhat,s) e^{-n\hbar\rd/\rd s},    \\
  \Lhat =& \sum_{n=1}^\infty
         \uhat_n(\hbar,z,\zhat,s) e^{n\hbar\rd/\rd s}    \tag\eq \\
\endalign
$$
and $B_n$ and $\Bhat_n$ are given by
$$
   B_n = (L^n)_{\ge 0}, \quad
   \Bhat_n = (\Lhat^{-n})_{\le -1}.                \tag\eq
$$
Here $(\quad)_{\ge 0}$ and $(\quad)_{\le -1}$ denote the projection
onto a linear combination of $e^{n\hbar\rd/\rd s}$ with
$n\ge 0$ and $\le -1$ respectively.
\foot{%
  These notations are quite different from earlier papers
  [\UeTa]. In particular, $L$ and $\Lhat$ correspond to $L$ and $M$
  therein, whereas $M$ and $\Mhat$ introduced here have no counterpart.
  Furthermore, $W$ and $\What$ introduced below correspond to
  $\What^{(\infty)}$ and $\What^{(0)}$ in those papers, whereas
  $\Psi$ and $\Psihat$ to $W^{(\infty)}$ and $W^{(0)}$ therein.
  We would like to apologize for this notational inconsistency.}
This Lax representation, too, can be extended into a larger system.
Besides the above Lax equations, the extended Lax representation
contains the second set of Lax equations
$$
\align
  \hbar \frac{\rd M}{\rd z_n} = [ B_n, M ], \quad&
  \hbar \frac{\rd M}{\rd \zhat_n} = [\Bhat_n, M ],
                                                                  \\
  \hbar \frac{\rd \Mhat}{\rd z_n} = [ B_n, \Mhat ], \quad&
  \hbar \frac{\rd \Mhat}{\rd \zhat_n} = [ \Bhat_n, \Mhat ]
  \quad n = 1,2,\ldots,                                   \tag\eq \\
\endalign
$$
and the canonical commutation relations
$$
  [ L, M ] = \hbar L, \quad  [ \Lhat, \Mhat ] = \hbar \Lhat  \tag\eq
$$
for a second set of difference operators
$$
\align
  M =& \sum_{n=1}^\infty n z_n L^n + s
          +\sum_{n=1}^\infty v_n(\hbar,z,\zhat,s)L^{-n},
                                                                   \\
  \Mhat =& -\sum_{n=1}^\infty n \zhat_n \Lhat^{-n} + s
          +\sum_{n=1}^\infty \vhat_n(\hbar,z,\zhat,s)\Lhat^n.
                                                           \tag\eq \\
\endalign
$$
These equations obviously bear a close resemblance to ordinary
quantum mechanics.
\foot{
  To be consistent with ordinary quantum mechanics, $\hbar$ in
  these formulas should be replaced by $i\hbar$.}

The dispersionless hierarchy emerges from the ordinary hierarchy
as quasi-classical limit as follows. We assume smooth asymptotic
behavior of the coefficients of the Lax operators as $\hbar \to 0$:
$$
\align
  u_n(\hbar,z,\zhat,s) =& u^{(0)}_n(z,\zhat,s) + O(\hbar),  \\
  v_n(\hbar,z,\zhat,s) =& v^{(0)}_n(z,\zhat,s) + O(\hbar),
                                                            \\
  \uhat_n(\hbar,z,\zhat,s) =& \uhat^{(0)}_n(z,\zhat,s) + O(\hbar),  \\
  \vhat_n(\hbar,z,\zhat,s) =& \vhat^{(0)}_n(z,\zhat,s) + O(\hbar).
                                                           \tag\eq  \\
\endalign
$$
One can then define Laurent series $\calL$, $\calM$, $\calLhat$ and
$\calMhat$ from these coefficients as
$$
   \calL = p + \sum_{n=1}^\infty u^{(0)}_{n+1}(z,\zhat,s) p^{-n},
   \quad \text{etc.,}
                                                          \tag\eq
$$
and Laurent polynomials $\calB_n(p)$ and $\calBhat_n(p)$,
similarly, from the coefficients
$$
\align
  b_{n,i}(\hbar,z,\zhat,s)
  =& b_{n,i}^{(0)}(z,\zhat,s) + O(\hbar),                          \\
  \bhat_{n,i}(\hbar,z,\zhat,s)
  =& \bhat_{n,i}^{(0)}(z,\zhat,s) + O(\hbar)
                                                          \tag\eq  \\
\endalign
$$
of the difference operators
$$
\align
  B_n(p)
  =& e^{n\hbar\rd/\rd s}
  + \sum_{i=0}^{n-2} b_{n,i}(\hbar,z,\zhat,s) e^{i\hbar\rd/\rd s}, \\
  \Bhat_n(p)
  =& \sum_{i=1}^n \bhat_{n,i}(\hbar,z,\zhat,s) e^{-i\hbar\rd/\rd s}
                                                       \tag\eq     \\
\endalign
$$
as
$$
\align
  \calB_n(p) =& p^n + \sum_{i=0}^{n-2} b^{(0)}_{n,i}(z,\zhat,s) p^i, \\
  \calBhat_n(p) =& \sum_{i=1}^n \bhat^{(0)}_{n,i}(z,\zhat,s) p^{-i}.
                                                         \tag\eq     \\
\endalign
$$
In quasi-classical ($\hbar \to 0$) limit, commutators of
difference operators turn into Poisson brackets as:
$$
  [ e^{\hbar\rd/\rd s}, s] = \hbar e^{\hbar\rd/\rd s} \quad
  \longrightarrow \quad
  \{ p, s \} = s.                                         \tag\eq
$$
The Lax equations of $L$, $M$, $\Lhat$ and $\Mhat$ can thereby be
reduced to the dispersionless Lax equations of $\calL$, $\calM$,
$\calLhat$ and $\calMhat$.

\section{Dressing operators, Baker-Akhiezer functions and tau function}

\noindent
The notions of dressing operators, Baker-Akhiezer functions and
tau function [\UeTa] can be reformulated so as to fit into
the above setting.

The dressing operators are now given by difference operators of the form
$$
\align
  W =& 1 + \sum_{n=1}^\infty
               w_n(\hbar,z,\zhat,s)e^{-n\hbar\rd/\rd s},
                                                                  \\
  \What =& \sum_{n=0}^\infty
           \what_n(\hbar,z,\zhat,s)e^{n\hbar\rd/\rd s}.   \tag\eq
                                                                  \\
\endalign
$$
The Lax operators are then given by the ``dressing" relations
$$
\align
  & L = W e^{\hbar\rd/\rd s} W^{-1}, \quad
    \Lhat = \What e^{\hbar\rd/\rd s} \What^{-1},
                                                               \\
  & M = W \left(
         \sum_{n=1}^\infty n z_n e^{n\hbar\rd/\rd s} + s \right)
      W^{-1},
                                                               \\
  & \Mhat = \What \left(
           -\sum_{n=1}^\infty n \zhat_n e^{-n\hbar\rd/\rd s}+ s \right)
           \What^{-1}.                            \tag\eqname\DressRel
                                                               \\
\endalign
$$
The Lax equations can be converted into the evolution equations
$$
\align
  \hbar \frac{\rd W}{\rd z_n} = B_n W - W e^{n\hbar\rd/\rd s}, \quad&
  \hbar \frac{\rd W}{\rd \zhat_n} = \Bhat_n W,
                                                                   \\
  \hbar \frac{\rd \What}{\rd z_n} = B_n \What,                 \quad&
  \hbar \frac{\rd \What}{\rd \zhat_n} =
                  \Bhat_n \What - \What e^{-n\hbar\rd/\rd s}.
                                                   \tag\eqname\FlowOfW
                                                                   \\
\endalign
$$
The coefficients $w_n$ and $\what_n$, unlike $u_n$ etc., are singular
as $\hbar \to 0$, as we shall see in the following  analysis of
Baker-Akhiezer functions.

Baker-Akhiezer functions are given by (formal) Laurent series of
a ``spectral parameter" $\lambda$:
$$
\align
  & \Psi = \left( 1 + \sum_{n=1}^\infty
              w_n(\hbar,z,\zhat,s)\lambda^{-n} \right)
           \exp \hbar^{-1}[ z(\lambda) + s\log\lambda],
                                                                    \\
  & \Psihat = \left( \sum_{n=0}^\infty
                 \what_n(\hbar,z,\zhat,s)\lambda^n \right)
              \exp \hbar^{-1}[ \zhat(\lambda^{-1}) +s\log\lambda],
                                                         \tag\eq    \\
\endalign
$$
where
$$
   z(\lambda) = \sum_{n=1}^\infty z_n \lambda^n, \quad
    \zhat(\lambda^{-1}) = \sum_{n=1}^\infty \zhat_n \lambda^{-n}.
                                                         \tag\eq
$$
Dressing relations (\DressRel) can now be transformed into
linear equations of $\Psi$ and $\Psihat$:
$$
\align
  \lambda \Psi = L \Psi, \quad &
  \hbar \lambda \frac{\rd \Psi}{\rd \lambda} = M \Psi,           \\
  \lambda \Psihat = \Lhat \Psihat, \quad &
  \hbar \lambda \frac{\rd \Psihat}{\rd \lambda} = \Mhat \Psihat.
                                                        \tag\eq  \\
\endalign
$$
In particular, the coefficients $v_n$ and $\vhat_n$ of
$M$ and $\Mhat$ can be read off from logarithmic derivatives
of $\Psi$ and $\Psihat$:
$$
\align
  \hbar\lambda \frac{\rd \log\Psi}{\rd \lambda}
  =& \sum_{n=1}^\infty n z_n \lambda^n + s
     + \sum_{n=1}^\infty v_n(\hbar,z,\zhat,s)\lambda^{-n},
                                                                 \\
  \hbar\lambda \frac{\rd \log\Psihat}{\rd \lambda}
  =& -\sum_{n=1}^\infty n \zhat_n \lambda^{-n} + s
     + \sum_{n=1}^\infty \vhat_n(\hbar,z,\zhat,s)\lambda^n.
                                                        \tag\eq
                                                                 \\
\endalign
$$
Evolution equations (\FlowOfW) of the dressing operators, too,
can be converted into linear equations of $\Psi$ and $\Psihat$:
$$
\align
  \hbar \frac{\rd\Psi}{\rd z_n} = B_n \Psi,              \quad &
  \hbar \frac{\rd\Psi}{\rd \zhat} = \Bhat_n \Psi         \\
  \hbar \frac{\rd\Psihat}{\rd z_n} = B_n \Psihat         \quad &
  \hbar \frac{\rd\Psihat}{\rd \zhat} = \Bhat_n \Psihat
                                                 \tag\eq \\
\endalign
$$
These equations resemble time-dependent Schr\"odinger equations
in ordinary quantum mechanics.  This implies that $\Psi$ and
$\Psihat$ takes a WKB asymptotic form as $\hbar \to 0$:
$$
\align
  \Psi =& \exp\left[ \hbar^{-1} S(z,\zhat,s,\lambda)
                                     + O(\hbar^0)\right],      \\
  \Psihat =& \exp\left[ \hbar^{-1} \Shat(z,\zhat,s,\lambda)
                                     + O(\hbar^0)\right],
                                                   \tag\eqname\WKB
                                                               \\
\endalign
$$
where $S(z,\zhat,s,\lambda)$ and $\Shat(z,\zhat,s,\lambda)$
have Laurent expansion of the form
$$
\align
  S(z,\zhat,s,\lambda) =& z(\lambda) + s\log\lambda
         + \sum_{n=1}^\infty S_n(z,\zhat,s) \lambda^{-n}, \\
  \Shat(z,\zhat,s,\lambda) =& \zhat(\lambda^{-1}) + s\log\lambda
         + \sum_{n=0}^\infty \Shat_n(z,\zhat,s) \lambda^n.
                                                           \tag\eq
                                                                   \\
\endalign
$$
In particular, $w_n$ and $\what_n$ are singular as $\hbar \to 0$.

The tau function, too, exhibits characteristic singular behavior
as $\hbar \to 0$.  In the presence of $\hbar$, we define the tau
function  $\tau(\hbar,z,\zhat,s)$ as a function that reproduces
the Baker-Akhiezer functions as
$$
\align
  & \dfrac{ \tau(\hbar,z-\hbar\epsilon(\lambda^{-1}),\zhat,s) }
        { \tau(\hbar,z,\zhat,s) }
  \exp \hbar^{-1}[ z(\lambda) + s\log\lambda ]
  = \Psi(\hbar,z,\zhat,\lambda),
                                                                 \\
  & \dfrac{ \tau(\hbar,z,\zhat-\hbar\epsilon(\lambda),s+\hbar) }
        { \tau(\hbar,z,\zhat,s) }
  \exp \hbar^{-1}[ \zhat(\lambda^{-1}) + s\log\lambda ]
  = \Psihat(\hbar,z,\zhat,\lambda),
                                            \tag\eqname\TauDef \\
\endalign
$$
where
$$
    \epsilon(\lambda) = \left( \lambda, \frac{\lambda^2}{2},
        \ldots, \frac{\lambda^n}{n},\ldots \right).   \tag\eq
$$
In the case of $\hbar = 1$, this reduces to the ordinary definition.
Taking the logarithm of (\TauDef) and comparing them with
the WKB asymptotic form (\WKB) of $\Psi$ and $\Psihat$,
one can easily find that $\log\tau(\hbar,z,\zhat,s)$
should behave as
$$
  \log\tau(\hbar,z,\zhat,s) = \hbar^{-2} F(z,\zhat,s) + O(\hbar^{-1})
  \quad (\hbar \to 0)                               \tag\eqname\TauAsymp
$$
with an appropriate scaling function $F(z,\zhat,s)$.
The Laurent coefficients $S_n$ and $\Shat_n$ of
$S(z,\zhat,s,\lambda)$ and $\Shat(z,\zhat,s,\lambda)$
can be written
$$
  S_n = -\frac{1}{n} \frac{\rd F}{\rd z_n}, \quad
  \Shat_n = -\frac{1}{n} \frac{\rd F}{\rd \zhat_n}, \quad
  \Shat_0 = \frac{\rd F}{\rd s}.   \tag\eqname\SandF
$$
This implies that $F(z,\zhat)$ is nothing but the logarithm
of the tau function of the dispersionless Toda hierarchy [\TaTadToda]:
$$
  F = \log \tau_{\text{dToda}}                        \tag\eq
$$
The function $F$ may be called the ``free energy"  in analogy
with matrix models of two dimensional quantum gravity [\TaTaQCKP].

\section{Hamilton-Jacobi equations and Legendre transformation}

\noindent
The ``phase functions" $S$ and $\Shat$ satisfy a set of
Hamilton-Jacobi equations, which turn out to reproduce the
Lax formalism of the dispersionless Toda hierarchy after a Legendre
transformation.  To see this, let us gather up the
Hamilton-Jacobi equations into 1-form equations:
$$
\align
  dS(z,\zhat,s,\lambda)
  =& \calM(\lambda) \frac{d\lambda}{\lambda}
   +\frac{\rd S}{\rd s}ds
   +\sum_{n=1}^\infty
        \calB_n\left( e^{\rd S/\rd s} \right) dz_n
   +\sum_{n=1}^\infty
        \calBhat_n\left( e^{\rd S/\rd s} \right)d\zhat_n,
                                                                    \\
  d\Shat(z,\zhat,s,\lambda)
  =& \calMhat(\lambda) \frac{d\lambda}{\lambda}
   +\frac{\rd \Shat}{\rd s}ds
   +\sum_{n=1}^\infty
        \calB_n\left( e^{\rd \Shat/\rd s} \right) dz_n
   +\sum_{n=1}^\infty
        \calBhat_n\left( e^{\rd \Shat/\rd s} \right)d\zhat_n,       \\
   &                                                       \tag\eq  \\
\endalign
$$
where
$$
\align
   \calM(\lambda) =& \sum_{n=1}^\infty n z_n \lambda^n +s
       +\sum_{n=1}^\infty v^{(0)}_n(z,\zhat,s) \lambda^{-n},
                                                                    \\
   \calMhat(\lambda) =& -\sum_{n=1}^\infty n \zhat_n \lambda^{-n} +s
       +\sum_{n=1}^\infty \vhat^{(0)}_n(z,\zhat,s) \lambda^n.
                                                         \tag\eq
                                                                    \\
\endalign
$$
Exterior differentiation of these equations give a 2-form equation of
the form
$$
\align
  \frac{d\lambda}{\lambda} \wedge d\calM(\lambda)
  =& d\left( \frac{\rd S}{\rd s} \right) \wedge ds
    +\sum_{n=1}^\infty
        d\calB_n\left(e^{\rd S/\rd s}\right) \wedge dz_n
    +\sum_{n=1}^\infty
        d\calBhat_n\left(e^{\rd S/\rd s}\right) \wedge d\zhat_n   \\
                                                        & \tag\eq \\
\endalign
$$
and a similar 2-form equation including $\calMhat(\lambda)$ and
$\Shat$ in place of $\calM(\lambda)$ and $S$.
We now distinguish between the two $\lambda$'s in
$S$ and $\Shat$ as $S(z,\zhat,s,\lambda)$ and
$\Shat(z,\zhat,s,\lambdahat)$, and solve the equations
$$
    \exp \rd S(z,\zhat,s,\lambda) / \rd s
  = \exp \rd \Shat(z,\zhat,s,\lambdahat) / \rd s
  = p                                                  \tag\eq
$$
with respect to $\lambda$ and $\lambdahat$.  Obviously this is
a kind of Legendre transformation. Let us write the solutions
$$
    \lambda = \calL(z,\zhat,s,p), \quad
    \lambdahat = \calLhat(z,\zhat,s,p)                \tag\eq
$$
and define
$$
   \calM = \calM(\lambda)|_{\lambda=\calL(z,\zhat,s,p)}, \quad
   \calMhat = \calMhat(\lambdahat)|_{\lambdahat=\calLhat(z,\zhat,s,p)}.
                                                      \tag\eq
$$
Then the above 2-form equations coincide with (\TwoForm), hence
these $\calL$, $\calM$, $\calLhat$ and $\calMhat$ indeed satisfy
the dispersionless Toda hierarchy.

\section{W-infinity symmetries}

\noindent
We now turn to the issue of W-infinity symmetries. W-infinity
symmetries of the Toda hierarchy can be formulated in two different
ways, i.e., in bosonic and fermionic languages. For the analysis
of quasi-classical limit, the bosonic language is more convenient.

The bosonic description is based on the so called
``vertex operators"
[\REF\DJKM{
  Date, E., Kashiwara, M., Jimbo, M., and Miwa, T.,
  Transformation groups for soliton equations,
  in: {\it Nonlinear Integrable Systems ---
  Classical Theory and Quantum Theory}
  (World Scientific, Singapore, 1983).}
\DJKM].
Actually, the Toda hierarchy has two copies of
$W_{1+\infty}$ symmetries, which mutually commute.
They are realized by the infinitesimal action
$\tau \to \tau + \epsilon Z\tau$,
$\tau \to \tau + \epsilon \Zhat \tau$
of the vertex operators
$$
\align
  Z(\hbar,\lambdatilde,\lambda) =&
  \dfrac{ \exp\left(\hbar^{-1}[z(\lambdatilde)-z(\lambda)]\right)
          (\lambdatilde/\lambda)^{s/\hbar}
          \exp\left(\hbar[-\rdtilde_z(\lambdatilde^{-1})
                          +\rdtilde_z(\lambda^{-1})]\right)
          - 1}
        {\lambdatilde - \lambda},
                                                                      \\
  \Zhat(\hbar,\lambdatilde,\lambda) =&
  \dfrac{ \exp\left(\hbar^{-1}[\zhat(\lambdatilde^{-1})
                               -\zhat(\lambda^{-1})]\right)
          (\lambdatilde/\lambda)^{s/\hbar}
          \exp\left(\hbar[-\rdtilde_\zhat(\lambdatilde)
                          +\rdtilde_\zhat(\lambda)]\right)
          - 1}
        {\lambdatilde^{-1} - \lambda^{-1}},                           \\
   &                                                        \tag\eq   \\
\endalign
$$
where
$$
  \rdtilde_z(\lambda^{-1}) = \sum_{n=1}^\infty
        \frac{\lambda^{-n}}{n} \frac{\rd}{\rd z_n},  \quad
  \rdtilde_\zhat(\lambda) = \sum_{n=1}^\infty
        \frac{\lambda^n}{n} \frac{\rd}{\rd \zhat_n}.   \tag\eq
$$
If one expands these two-parameter families of symmetries
into Fourier modes along the double loop
$|\lambdatilde| = |\lambda| = \text{const.}$, the outcome are
two copies of gl($\infty$) symmetries [\DJKM]. If one first
expands these vertex operators into Taylor series along the
diagonal $\lambdatilde = \lambda$,
$$
  Z(\hbar,\lambdatilde,\lambda) = \sum_{\ell=1}^\infty
      \dfrac{(\lambdatilde-\lambda)^{\ell-1} }
            { (\ell-1)! }
      W^{(\ell)}(\hbar,\lambda),
  \quad \text{etc.},                                  \tag\eq
$$
and further into Fourier modes along the loop
$|\lambda|=\text{const.}$,
$$
  W^{(\ell)}(\hbar,\lambda) = \sum_{n=-\infty}^\infty
       W^{(\ell)}_n(\hbar) \lambda^{-n-\ell},
  \quad  \text{etc.},                                 \tag\eq
$$
the coefficients $W^{(\ell)}_n(\hbar)$ and $\What^{(\ell)}_n(\hbar)$
($n \in \bfZ$, $\ell \ge 1$) become generators of $W_{1+\infty}$
symmetries. These symmetry generators are differential operators
of finite order in $z$ and $\zhat$ respectively, and differ from
the ordinary ($\hbar = 1$) definition [\DJKM] by the simple rescaling
$$
\align
  z_n \to \hbar^{-1} z_n, \quad &
  \frac{\rd}{\rd z_n} \to \hbar \frac{\rd}{\rd z_n},        \\
  \zhat_n \to \hbar^{-1} \zhat_n, \quad &
  \frac{\rd}{\rd \zhat_n} \to \hbar \frac{\rd}{\rd \zhat_n}.
                                                    \tag\eq \\
\endalign
$$
We have thus essentially the same $W_{1+\infty}$ symmetries as
the KP hierarchy, but now in duplicate.

Let us show how to contract these $W_{1+\infty}$ symmetries into
$w_{1+\infty}$ symmetries of the dispersionless hierarchy. The
essence is the same as in the case of the KP hierarchy [\TaTaQCKP].
First, with the aid of basic relations (\TauAsymp) and (\SandF),
we write the action of $W^{(\ell)}(\hbar,\lambda)$ on the tau function
in terms of $\calM(\lambda)$:
$$
\align
  \dfrac{ W^{(\ell)}(\hbar,\lambda) \tau(\hbar,z,\zhat,s) }
        { \tau(\hbar,z,\zhat,s) }
  &= \left.
     \frac{1}{\ell} \left(\frac{\rd}{\rd \lambdatilde}\right)^\ell
     \exp \hbar^{-1}[ S(z,\zhat,s,\lambdatilde)
                     -S(z,\zhat,s,\lambda) +O(\hbar)]
     \right|_{\lambdatilde=\lambda}                                   \\
  &= \frac{\hbar^{-\ell}}{\ell} \left[
         \left( \frac{\rd S(z,\zhat,s,\lambda)}{\rd \lambda} \right)^\ell
         +O(\hbar) \right]                                            \\
  &= \frac{\hbar^{-\ell}}{\ell} \left[
         \left( \calM(\lambda)\lambda^{-1} \right)^\ell
         +O(\hbar) \right].                                 \tag\eq   \\
\endalign
$$
We then pick out the most singular term ($\propto \hbar^{-\ell}$)
as $\hbar \to 0$, and consider it as defining a $\lambda$-dependent
infinitesimal transformation of $F$,
$F \to F + \epsilon w^{(\ell)}(\lambda)F$.
Its Fourier modes $w^{(\ell)}_n F$ are given by
$$
\align
  w^{(\ell)}_n F
  =& -\Res_{\lambda=\infty}
        \frac{1}{\ell} \left( \calM(\lambda)\lambda^{-1} \right)^\ell
        \lambda^{n+\ell-1} d\lambda                                    \\
  =& -\Res_{\lambda=\infty}
        \frac{1}{\ell} \calM^\ell \calL^n d\log\calL
                                                             \tag\eq   \\
\endalign
$$
where $\Res_{\lambda=\infty}$ denotes the residue at $\lambda=\infty$,
$$
  \Res_{\lambda=\infty} \lambda^n d\lambda = - \delta_{n,-1}.  \tag\eq
$$
These $w^{(\ell)}_n$ coincide with one half of the $w_{1+\infty}$
symmetries that have been constructed by a different method [\TaTadToda].
Another half can be obtained from $\What^{(\ell)}_n(\hbar)$
in the same way.

In the fermionic language, the vertex operators correspond to
fermion bilinear forms. Let $\psi(\lambda)$ and $\psi^*(\lambda)$
be the Date-Jimbo-Kashiwara-Miwa free fermion fields [\DJKM]
$$
  \psi(\lambda) = \sum_{n=-\infty}^\infty \psi_n \lambda^n, \quad
  \psi^*(\lambda) = \sum_{n=-\infty}^\infty \psi^*_n \lambda^{-n-1}
                                                            \tag\eq
$$
with anti-commutation relations
$$
  [\psi_i,\psi_j]_{+} = [\psi^*_i,\psi^*_j]_{+} = 0, \quad
  [\psi_i,\psi^*_j]_{+} = \delta_{ij},                      \tag\eq
$$
and $<n\mid $ and $\mid n>$, $n \in \bfZ$, be the ground states in
the charge-$n$ sector of the Fock space,
$$
\align
  & \psi_n \mid 0>   = 0  \quad (n \le -1), \quad
    \psi^*_n \mid 0> = 0  \quad (n \ge 0),                        \\
  & <0\mid  \psi_n   = 0  \quad (n \ge 0),  \quad
    <0\mid  \psi^*_n = 0  \quad (n \le -1).              \tag\eq  \\
\endalign
$$
A generic expression of the tau function is given by
[\REF\TodaTau{
  Jimbo, M., and Miwa, T.,
  Solitons and infinite dimensional Lie algebras,
  Publ. RIMS, Kyoto Univ., 19 (1983), 943-1001. \nextline
  Takebe, T.,
  Representation theoretical meaning of the initial value problem
  for the Toda lattice hierarchy: I,
  Lett. Math. Phys. 21 (1991), 77-84;
  ditto II,
  Publ. RIMS, Kyoto Univ., 27 (1991), 491-503.}
\TodaTau]
$$
  \tau(\hbar,z,\zhat,s)
  = <\hbar^{-1}s  \mid  e^{H(z)/\hbar} g(\hbar)
                  e^{-\Hhat(\zhat)/\hbar} \mid \hbar^{-1}s >,   \tag\eq
$$
where $g(\hbar)$ is an appropriate $\hbar$-dependent Clifford
operator, and $H(z)$ and $\Hhat(\zhat)$ are generators of time
evolutions,
$$
\align
  & H(z) = \sum_{n=1}^\infty z_n H_n, \quad
    \Hhat(\zhat) = \sum_{n=1}^\infty \zhat_n H_{-n},       \\
  & H_n = \sum_{m=-\infty}^\infty : \psi_m \psi^*_{m+n}:
                                                 \tag\eq   \\
\endalign
$$
normal ordered with respect to $<0\mid $ and $\mid 0>$. Actually,
the Clifford operator $g(\hbar)$ takes the same form as in the
case of the KP hierarchy [\TaTaQCKP]:
$$
  g(\hbar) = \exp \hbar^{-1} \oint
      :A\left(\lambda,\hbar\frac{\rd}{\rd \lambda}\right)
      \psi(\lambda) \cdot \psi^*(\lambda): \frac{d\lambda}{2\pi i},
                                                  \tag\eq
$$
where $A$ is a linear differential operator, and
``$A\psi \cdot \psi^*$" means $A\psi$ times $\psi^*$.
The action of $Z(\hbar,\lambdatilde,\lambda)$ and
$\Zhat(\hbar,\lambdatilde,\lambda)$ is realized by
insertion of a fermion bilinear form:
$$
\align
  & Z(\hbar,\lambdatilde,\lambda) \tau(\hbar,z,\zhat,s)
    = <\hbar^{-1}s \mid  e^{H(z)/\hbar}
           \psi(\lambdatilde)\psi^*(\lambda) g(\hbar)
           e^{-\Hhat(\zhat)/\hbar} \mid \hbar^{-1}s >,          \\
  & \Zhat(\hbar,\lambdatilde,\lambda) \tau(\hbar,z,\zhat,s)
    = <\hbar^{-1}s \mid  e^{H(z)/\hbar}
           g(\hbar) \psi(\lambdatilde)\psi^*(\lambda)
           e^{-\Hhat(\zhat)/\hbar} \mid \hbar^{-1}s >.         \tag\eq   \\
\endalign
$$
The bosonic and fermionic representations are thus connected.
For the moment, the bosonic language looks more preferable,
because the fermionic representation is valid only for
discrete values $ s \in \hbar\bfZ $.

\section{Conclusion}

We have thus extended our previous results on the quasi-classical
limit of the KP hierarchy to the Toda hierarchy. The Planck constant
$\hbar$ now enters into the Lax formalism as the spacing unit of
difference operators.  The notions of dressing operators,
Baker-Akhiezer functions and tau function are redefined so as
to fit into the new formulation. We have used two copies of the
Date-Jimbo-Kashiwara-Miwa vertex operators to calculate the action
of $W_{1+\infty}$ symmetries on the tau function, which exibits
singular behavior as $\hbar \to 0$. The most singular terms therein
turn out to give $w_{1+\infty}$ symmetries of the dispersionless
Toda hierarchy.

Hopefully, these results will further be exntended to multi-component
KP and Toda hierarchies. The so called Whitham hierarchies
[\REF\DuKr{
  Dubrovin, B.A.,
  Hamiltonian formalism of Whitham-type hierarchies
  and topological Landau-Ginsburg models,
  Commun. Math. Phys. 145 (1992), 195-207.\nextline
  Krichever, I.M.,
  The $\tau$-function of the universal Whitham hierarchy,
  matrix models and topological field theories,
  LPTENS-92/18 (May, 1992).}
\DuKr]
will then emerge as quasi-classical limit.

This work is supported in part by the Grant-in-Aid for
Scientific Research, the Ministry of Education, Science
and Culture, Japan.

\refout
\bye